# Evolution Of Gravitational Potential In The Quasilinear And Nonlinear Regimes

J.S.Bagla[1] and T.Padmanabhan[2]

*Inter-University Centre for Astronomy and Astrophysics*
*Post Bag 4, Ganeshkhind,*
*Pune - 411 007, INDIA.*

**Abstract**

We study the evolution of the power spectrum of gravitational potential during the nonlinear clustering in an $\Omega = 1$ matter dominated phase. N-body simulations suggest that the potential does not evolve in time even in the quasilinear phase for an $n = -1$ power spectrum. For $n = -2$, the potential evolves in the quasilinear phase but not in the extreme nonlinear phase. Becauase of these facts, the evolution of the gravitational potential in spectra like CDM is less than what would have been expected naively. We discuss a class of CDM like models in which such an interesting conspiracy between the amplitude and local index occurs.

email : [1] jasjeet@iucaa.ernet.in    [2] paddy@iucaa.ernet.in

## 1. Introduction

The driving force behind the formation of large scale structures in the universe is the gravitational field produced by inhomogeneities. Overdense regions accrete matter at the expense of underdense regions allowing inhomogeneities in the universe to grow. At scales with $L \ll H^{-1}$ where $H = (\dot{a}/a)$ it is essentially this process which allows small initial inhomogeneities to grow into structures like galaxies, clusters etc. Further, observations suggest that the material content of the universe is dominated by dark matter which may be considered to be made of collisionless elementary particles. In that case, the gravitational force is dominated by these particles and, to first approximation, we can ignore complications arising from baryonic physics. The evolution of inhomogeneities is then governed purely by the gravitational force.

It is, therefore, intersting to ask the question: How does the gravitaional potential in the universe evolve as structures form ? In addition to the inherent academic interest, this question has two other facets to it.

To begin with, we note that the gravitational potential $\varphi(a, \mathbf{x})$ due to inhomogeneities is governed by the Poisson equation

$$\nabla^2 \varphi = 4\pi G \rho_b a^2 \delta = \frac{3}{2} H_0^2 \Omega_0 \left( \frac{\delta}{a} \right) \tag{1}$$

where $\delta(a, \mathbf{x})$ is the density contrast and $\rho_b$ is the background density. [ This equation is valid in the limit of $L \ll H^{-1}$ where $L$ is the scale we are interested in. Such a description is quite adequate for the study of nonlinear structure formation.] When the perturbations are small and the linear theory is applicable, $\delta \propto a$ if $\Omega_0 = 1$ and the universe is matter dominated. [If $\Omega_0 \neq 1$, the gravitational potential undergoes slow evolution during the linear regime that can be computed using linear theory. In this paper we shall concentrate on the $\Omega_0 = 1$ model.] Hence the nontrivial evolution of gravitational potential takes place only in the nonlinear phase. Because of this reason the study of growth of gravitational potential offers a direct diagnostic of the nonlinear dynamics.

The second aspect is the following: Sometime back the authors have suggested an approximation scheme for the study of gravitational dynamics called "frozen potential approximation" [Bagla and Padmanabhan, 1994; this was also suggested independently by Brainerd et al, 1993]. Comparison of this approximation scheme with N-body results shows that it works well in the quasilinear regime and provides a reasonably accurate information about velocities. In fact, for certain class of spectra this scheme works unexpectedly well, showing that the average effect of growth of gravitational potential is not as important as one would have naively imagined. *This requires an explanation.* In particular, one would like to know whether there exists a class of spectra for which gravitational potential does *not* evolve significantly even in the nonlinear regime. We shall address these questions in this paper.

In principle, the question of evolution of gravitational potential can be settled in a straightforward manner by running suitable N-body simulations. However, such an approach does not provide one with an intuitive understanding of the results. Because of this reason we shall follow a more indirect route in this paper which, it turns out, produces results that are equivalent to the N-body simulations (to the accuracy we want). This approach is based on the observation that numerical simulations suggest a simple relationship between the mean relative pair velocities of particles $v(a,x)$ and the mean correlation function $\bar{\xi}(a,x)$ at the same epoch and scale [Hamilton et al., 1991]. The mean correlation function $\bar{\xi}(a,x)$ is defined as

$$\bar{\xi}(a,x) = \frac{3}{x^3} \int_0^x \xi(a,y) y^2 dy \tag{2}$$

Let $h(a,x) \equiv [-v(a,x)/\dot{a}x]$ be the dimensionless relative pair velocity at the comoving scale $x$ at the epoch $a$. Then, following Nityananda and Padmanabhan [Nityananda and Padmanabhan, 1994] we postulate that $h(a,x)$ is a universal function of $\bar{\xi}(a,x)$ evaluated at the same epoch and scale; that is

$$h(a,x) = Q\left[\bar{\xi}(a,x)\right] \tag{3}$$

where the function $Q$ can be well approximated by

$$h(\bar{\xi}) = \begin{cases} \frac{2\bar{\xi}}{(3+\bar{\xi})} & \text{(for } \bar{\xi} \leq 38\text{)} \\ \left(\frac{\bar{\xi}}{560}\right)^{-0.23} & \text{(for } 38 \leq \bar{\xi} \leq 100\text{ )} \\ \exp\left(\frac{40}{\bar{\xi}}\right) & \text{(for } 100 \leq \bar{\xi}\text{)} \end{cases} \tag{4}$$

It can be shown that [Nityananda and Padmanabhan, 1994] this result allows one to express the true mean correlation function $\bar{\xi}(a,x)$ in terms of the correlation function in the linear theory $\bar{\xi}_L(a,l)$ as

$$\bar{\xi}_L(a,l) = \exp\left(\frac{2}{3} \int^{\bar{\xi}(a,x)} \frac{dq}{h(q)(1+q)}\right) \tag{5}$$

where $l = x\left[1 + \bar{\xi}(a,x)\right]^{1/3}$. For the functional form in (4) the relation between $\bar{\xi}(a,x)$ and $\bar{\xi}_L(a,l)$ can be well approximated by [Hamilton et al., 1991]

$$\bar{\xi}(a,x) = \frac{x + 0.358x^3 + 0.0236x^6}{1 + 0.0134x^3 + 0.00202x^{9/2}} \tag{6}$$

where $x = \bar{\xi}_L(a,l)$. In other words, one can find the nonlinear evolution of the correlation function and related quantities from the linear theory using the above ansatz. It is, however, possible to produce a still simpler fitting function which captures the essence of this formula: [Bagla and Padmanabhan, 1993]

$$\bar{\xi} = \begin{cases} \bar{\xi}_L & (\text{for } \bar{\xi}_L < 1.2, \bar{\xi} < 1.2) \\ 0.7\bar{\xi}_L^3 & (\text{for } 1.2 < \bar{\xi}_L < 6.5, 1.2 < \bar{\xi} < 195) \\ 11.7\bar{\xi}_L^{3/2} & (\text{for } 6.5 < \bar{\xi}_L, 195 < \bar{\xi}) \end{cases} \quad (7)$$

As we shall see, this simple form allows one to understand clearly the effects which are operating at different scales. It is apparent that evolution of density contrast is generically characterised by two different scales: At about $\bar{\xi}_L \approx 1$ the deviations from linear theory begin to manifest itself. The second scale occurs around $\bar{\xi}_L \approx 6$ [with $\bar{\xi} \approx 200$] when we expect virialised structures to form and separate out from the overall dynamics. These two scales are characterised by the different slopes in (7). For the sake of definiteness we shall call the three regimes linear, quasilinear and nonlinear.

Before proceeding further, we would like to make some comments regarding the accuracy of the above ansatz. This relation was first obtained by Hamilton et al., based on N-body simulation data of Efsthathiou et al [ Hamilton et al., 1991; Efstathiou et al., 1988]. In the original paper, it was claimed that the relationship is reasonably good and valid at all epochs and for all spectra. Recently, this relation has been tested by more accurate simulations with larger dynamic range [Padmanabhan et al., 1995]. The results of this paper suggests that the "universality" is only approximately valid. The form and asymptotic limit of $h$ has a weak spectrum dependence and the asymptotic value is lower for spectra with more small scale power. However, this analysis also shows that the deviations from the universality are small and are at about 20% level. To this level of accuracy, one can use the above ansatz. The accuracy can, of course, be easily improved by fitting a more accurate [spectrum dependent] curve to the results. We shall not do so, since the key idea of this paper is only to develop a simple physical picture of growth of the gravitational potential.

## 2. The Gravitational Potential and Force

The results in (6) and (7) relate the mean correlation funtion for density perturbations in the linear and nonlinear regimes. To obtain the information about the gravitational potential from this result, the most direct approach for studying the evolution of gravitational potential is the following: The correlation function $C(a, \mathbf{x})$ for the dimensionless gravitational potential $\psi(a, \mathbf{x}) \equiv (2/3) H_0^{-2} \Omega_0^{-1} \varphi(a, \mathbf{x})$ defined as

$$C(a, \mathbf{x}) = \langle \psi(a, \mathbf{y} + \mathbf{x}) \psi(a, \mathbf{y}) \rangle_\mathbf{y} \tag{8}$$

is related to the matter correlation function $\xi(a, \mathbf{x})$ by the relation

$$\nabla^2 \nabla^2 C = \left[ \frac{1}{x^2} \frac{d}{dx} \left( x^2 \frac{d}{dx} \right) \right] \left[ \frac{1}{x^2} \frac{d}{dx} \left( x^2 \frac{dC(a,x)}{dx} \right) \right] = \frac{\xi(a,x)}{a^2} \tag{9}$$

where we have assumed that $C(a, \mathbf{x}) = C(a, x)$. The above equation follows immediately from the fact that the power spectra for density and potential are related by a factor $k^4$ in the fourier space. Given a correlation function in linear theory one can use (6) or (7) to obtain the nonlinear correlation function; integrating equation (9) we can find the evolution of correlations in gravitational potential. While integrating this equation, one can use the linear theory results at very large scales to obtain a unique solution. By fourier transforming the correlation function one can obtain the power spectrum of the gravitational potential.

It is, however, possible to simplify the above procedure by using the following fact : For spectra which vary sufficiently smoothly, the quantity $\Delta^2(a, k) \equiv \left[ k^3 P(k)(a, k)/2\pi^2 \right]$ at $k \simeq x^{-1}$ behaves in a manner very similar to $\bar{\xi}(a, x)$. So, instead of finding the correlation function of potential by integrating (9) and then obtaining the power spectra by fourier transforming, one can directly use the ansatz (6) in the fourier space to obtain the power spectrum of gravitational potential. [A similar relation has been suggested by Peacock and Dodds [Peacock and Dodds, 1994] for evolution of $\Delta^2$. As this relation is almost identical to (6) for $\Omega = 1$, we use (6) to obtain the nonlinear power spectrum. In their paper Peacock and Dodds make cautionary remarks about the applicability of (6) for spectra with $n < -1$. However the key results described in this paper depend only on the characteristic behaviour in the quasilinear and nonlinear regime as described in (7), which are seen in simulations of all spectra of physical interest.]

Such a simplification, of course, needs to be justified. We have used both methods and ascertained that the deviations are not significant in regimes of interest.

In addition to the gravitational potential we also investigate the evolution of the gradient of the potential, $\mathbf{g}(a, \mathbf{x}) \equiv -\nabla \psi$ which is more directly relevant in determining the dynamics. The correlation function for $\mathbf{g}(a, \mathbf{x})$ is a six-component object but we shall concentrate on the scalar

$$D(a, \mathbf{x}) \equiv \langle \mathbf{g}(a, \mathbf{y} + \mathbf{x}) . \mathbf{g}(a, \mathbf{y}) \rangle \tag{10}$$

The power spectrum associated with $D(a, \mathbf{x})$ is related to $P(a, k)$ by a $k^2$ factor. It is easy to see that $D(a, x)$ and $\bar{\xi}(a, x)$ are related by the equation

$$\frac{1}{x}\frac{dD(a,x)}{dx} = -\frac{1}{3}\frac{\bar{\xi}(a,x)}{a^2} \tag{11}$$

This can be integrated readily to obtain $D(a, x)$. A comparison with the linear correlation $D_L(a, x)$ is useful for estimating validity of quasilinear approximations at a given scale and epoch. Largest scale for which the change is above tolerance limit provides a natural criterion for validity of approximation schemes like frozen potential. Such a criterion can also be used to select a smoothing scale for truncated Zeldovich approximation. The comments made above regarding $C$ are also applicable to $D$.

## 3. Conclusions

We shall now present the results of our analysis. We first consider pure power law spectra for density with linear power spectrum $P(k) \propto k^n$ and $n = 1, 0, -1, -2$. For sake of uniformity all of these are normalised to give $\sigma_8 = 1$ at $a = 1$ with a gaussian window function. Initial power spectra for density and potential are $P_\delta(a, k) = Aa^2 k^n$; $P_\psi(a, k) = Ak^{n-4}$ with

$$A = \frac{4\pi^2 (8/h)^{(3+n)}}{\Gamma[(3+n)/2]}$$

Figures 1a and 1b give the evolution of $\Delta_\psi(k)$ and $\Delta_g(k)$ for $n = 1$ spectrum. Curves are plotted for five epochs $a = 0.1, 0.25, 0.5, 1, 2$. With our normalisation we expect the scale of $l = 8h^{-1}Mpc$ to go nonlinear around $a = 1$. We see that at the nonlinear end $\Delta_\psi(k)$ and $\Delta_g(k)$ change by a substantial amount, though the change here is less than the corresponding change in density correlations. Figure 1c gives the evolution of $\Delta(k)$ during the same epochs.

The results for power law spectra, of course, can be presented in a much more meaningful way. Our basic aim is to look at changes in potential in the nonlinear regime. For this, we should plot $\Delta_\psi$ and $\Delta_g$ as a function of $\Delta_L(k)$, the linearly evolved power spectrum for density fluctuations. Since there is no intrinsic scale in this problem, we expect this curve to evolve in a self-similar manner. Figures 1d and 1e show these plots for $n = 1$ spectrum. It can be shown below that $\Delta_\psi(k)$ is a time independent function of $\Delta_L(k)$ (see below), this implies that curves corresponding to different epochs will coincide, as shown in figure 1d.

The self similar scaling of these curves can be understood in the following manner. Consider $\Delta_\psi(k)$, it is related to the power spectrum of density fluctuations as

$$\Delta_\psi(k) = \frac{\Delta(k)}{a^2 k^4}. \tag{12}$$

We know that for power law spectra, power spectrum of density fluctuations must be a universal function of $[k/k_{nl}(a)]$ where $k_{nl}(a)$ is the scale that is going nonlinear at the epoch $a$ ; say, the scale at which $\Delta(k)$ is unity. This is also true of the linearly evolved power spectrum. If we can express $\Delta_\psi(k)$ in terms of a self similar function of $[k/k_{nl}(a)]$ and $a$, we would have obtained the relevant scaling. By rewriting (12) we obtain

$$\Delta_\psi(k) = \frac{1}{a^2 k_{nl}^4} \Delta(k/k_{nl}) \left(\frac{k_{nl}}{k}\right)^4 \propto a^{-2\frac{n-1}{n+3}} f(k/k_{nl}). \tag{13}$$

We have used the fact that $k_{nl} \propto a^{n+3}$ in deriving the scaling. Similar relation can also be derived for $\Delta_g(k)$,

$$\Delta_g(k) = \frac{1}{a^2 k_{nl}^2} \Delta(k/k_{nl}) \left(\frac{k_{nl}}{k}\right)^2 \propto a^{-2\frac{n+1}{n+3}} f(k/k_{nl}). \tag{14}$$

Here we notice an interesting fact : for $n = 1$, $\Delta_\psi(k)$ is a nonevolving function of $[k/k_{nl}(a)]$. Similarly for $n = -1$, $\Delta_g(k)$ is a time independent function of $[k/k_{nl}(a)]$.

Figures 2a and 2b give curves corresponding to figure 1d and 1e for $n = 0$ spectrum; since this contains the relevant information we have not plotted graphs corresponding to figures 1a, 1b and 1c in this case. The two scales, corresponding to quasilinear and nonlinear regime, are apparent in the graph.

We get our first surprise in the case of $n = -1$ spectrum, shown in figures 3a and 3b. We now see that the evolution follows linear result even in quasilinear regime, all the way up to $\Delta(k) \simeq 20$! In figure 3b, $\Delta_g(k)$ has been plotted for all the epochs mentioned above but as $\Delta_g(k)$ is a time independent function of $\Delta_L(k)$, the curves fall on top of each other. The corresponding evolution of density contrast is shown in figure 3c by plotting $\Delta(a, k)$ vs. $\Delta_L(a, k)$ at the same scale. It is seen that the nonlinear evolution follows the linear evolution closely upto $\Delta(a, k) \simeq 20$. [As far as the authors know, this fact has not been specifically noted previously in the literature.] This fact shows that $n = -1$ spectrum [corresponding to the isothermal density profile of $\rho \propto x^{-2}$ ] is rather special; for this spectrum, linear theory results have a validity which is beyond its legitimate domain.

It is, of course, possible to demonstrate this more explicitly using the fitting law in (7). For a powerlaw spectrum with index $n$, we can explicitly calculate the scaling relation with $a$ using this and we find that

$$\frac{\bar{\xi}(a,x)}{a^2 x^{-(n+3)}} \simeq \begin{cases} (.7)^{1/(n+4)} a^{-2\frac{n+1}{n+4}} x^{\frac{(n+1)(n+3)}{n+4}} & \text{(for } 1 \ll \bar{\xi} \leq 200 \text{ )} \\ (11.7)^{2/(n+5)} a^{-2\frac{n+2}{n+5}} x^{\frac{(n+2)(n+3)}{n+5}} & \text{(for } 200 \leq \bar{\xi} \text{ )} \end{cases} \tag{15}$$

This shows that if $n = -1$, then $\bar{\xi} \propto a^2$ even in the quasilinear regime. It is not clear how significant this result is and whether it is possible to provide a simple analytic argument to prove the same. [This issue is currently under investigation.]

Figures 4a and 4b give the results for $n = -2$. We see that there are deviations from linear theory in quasilinear regime but nonlinear evolution is similar to the linear one. This, of course, is a consequence of the scaling mentioned above. This result, taken in isolation, is not too interesting because: (i) the evolution in the quasilinear regime changes the amplitude, thereby making concrete predictions difficult and (ii) the result could be obtained trivially from the stable clustering argument applied to self-similar evolution. Figure 4c shows analogue of figure 3c for this spectrum.

But the above two results [for $n = -1$ and $-2$] taken together lead to an interesting "conspiracy" for a certain class of power spectra. To see this consider a power spectrum which has an index $-2$ in the nonlinear regime and $-1$ in the qusilinear regime. In the next instance, all the scales will grow in proportion to the expansion factor all the way from the strongly nonlinear regime to the linear regime. If we now smooth the spectrum and arrange its curvature such that the above condition is also [at least approximately] satisfied in the next instance, then the linear evolution will last longer. For such a spectrum, the gravitational potential will evolve very little.

Of course, this requires conspiracy between the slope and amplitude of the power spectrum at different scales. Surprisingly enough, CDM like spectra do have such a conspiracy built in to some extent. Figure 5c shows the local index of the COBE normalised CDM spectrum as a function of the density contrast, as the spectrum evolves. In order to maintain approximate constancy of the gravitational potential, the index should stay around $-2$ for $\Delta^2 \gtrsim 200$ and around $-1$ for $1 \lesssim \Delta^2 \lesssim 200$. Figures 5a, 5b and 5d give the corresponding curves for different values of $\Gamma$ for a CDM like spectra parametrised as

$$P(k) = \frac{Ak}{\left[\left(1 + bk + (ck)^{3/2} + (dk)^2\right)^{\nu}\right]^{2/\nu}} \tag{16}$$

Here $b$, $c$ and $d$ are functions of $\Gamma$. [We are following the parametrisation of Efstathiou et al., 1992]. It is usually claimed that $\Gamma = 0.2$ is a good fit to observations. We see that the conspiracy is stronger for smaller values of $\Gamma$ as the nonlinearities reached for these spectra are much lower than for spectra with high $\Gamma$. The evolution of C and D for these spectra is shown in figures 6a, 6b, 6c and 6d. In each figure we give the curves for $a = .1, .25, .5, 1, 2$. All the spectra are normalised to COBE. We see that the potential can be treated to be approximately constant [say, to 10 percent accuracy] up to scales of $2h^{-1}Mpc$ at $a = 1.0$ in the standard CDM. For the "best-fit" case of $\Gamma = 0.2$ there is no appreciable change in $\Delta_\psi(k)$ upto $a = 1$. However, if we compare models at the same level of nonlinearity at a scale of $1h^{-1}Mpc$ then spectra with higher value of $\Gamma$ fare better. This can be understood by plotting the power spectrum on the $n - \Delta$ plane [ See fig.5]. It is seen that for small $\Gamma$, the spectrum is flat and the physically relevant scales enter the nonlinear regime almost simulateneously, killing the conspiracy of indices and amplitude to some extent. However, for large $\Gamma$, power spectrum is steeper and the region with $n = -2$ is well into the nonlinear regime when scales with $n = -1$ are in the quasilinear regime. On the other hand, spectra with large $\Gamma$ have stronger nonlinearities as compared to models with small $\Gamma$, leading to smaller variations in $\psi(k)$. Another feature worth noting in Fig.5 is that for these spectra, a large range of scales develop a similar index, indicating power law correlations in that range of scales.

The results show that in hierarchical clustering models, two different factors combine together in deciding the evolution of the gravitational potential. Smaller scales reach the quasilinear and nonlinear phase at an earlier epoch and if their indices are appropriate, their growth rate becomes similar to that of linear evolution. On the other hand the curvature of the spectrum helps to bring in the right scale at the right amplitude. In real life, of course, all these are of only approximate validity; but it illustrates a very interesting dynamical aspect of the nonlinear gravitational clustering. By and large, the gravitational potential seems to be a rugged entity in the hierarchical models changing less than what might have been anticipated by naive arguments.

JSB would like to thank CSIR India for the Senior Research Fellowship.

## Figure Captions

**Fig. 1a** : Power spectrum for gravitational potential is plotted as a function of scale. Curves have been plotted for $P_\delta(k) = Ak^n$ with $n = 1$ at five epochs, $a = 0.1$, 0.25, 0.5, 1, 2. At the left end, lower curves correspond to later epochs.

**Fig. 1b** : Power spectrum for gravitational force. Spectrum and epochs are same as in Fig. 1a.

**Fig.1c** : Growth of Power spectrum of density fluctuations. Spectrum and epochs are same as in Fig. 1a.

**Fig.1d** : Power spectrum for potential is plotted against linearly extrapolated fluctuations in density. The dashed curve corresponds to the expected behaviour in linear theory. The dashed curve has been plotted for $a = 2$. Spectrum and epochs are same as in Fig. 1a.

**Fig. 1e** : Power spectrum for gravitational force. All other details are same as Fig. 1d.

**Fig. 2a** : Same as Fig. 1d but for $n = 0$.

**Fig. 2b** : Same as Fig. 1e but for $n = 0$.

**Fig. 3a** : Same as Fig. 1d but for $n = -1$.

**Fig. 3b** : Same as Fig. 1e but for $n = -1$.

**Fig. 3c** : Nonlinear power spectrum for density is plotted against linearly extrapolated power spectrum at the same scale for $n = -1$. Dashed line corresponds to linear behaviour. Notice the almost linear evolution of the nonlinear density contrast upto $\Delta(k) \simeq 20$.

**Fig. 4a** : Same as Fig. 1d but for $n = -2$.

**Fig. 4b** : Same as Fig. 1e but for $n = -2$.

**Fig. 4c** : Same as Fig. 3c but for $n = -2$. Here the nonlinear density contrast deviates from the linear density contrast in the quasilinear regime. However, in the nonlinear regime, both the curves are parallel to each other.

**Fig. 5a** : The power spectrum of density perturbations is plotted on the $n - \Delta$ plane for CDM spectrum with $\Gamma = 0.2$. Curves have been plotted for five epochs, $a = 0.1$, 0.25, 0.5, 1, 2.

**Fig. 5b** : Same as Fig. 5a but for model with $\Gamma = 0.35$.

**Fig. 5c** : Same as Fig. 5a but for model with $\Gamma = 0.5$.

**Fig. 5d** : Same as Fig. 5a but for model with $\Gamma = 0.7$.

**Fig. 6a** : Same as Fig.1a but for CDM, $\Gamma = 0.2$.

**Fig. 6b** : Same as Fig.1b but for CDM, $\Gamma = 0.2$.

**Fig. 6c** : Same as Fig.1a but for CDM, $\Gamma = 0.35$.

**Fig. 6d** : Same as Fig.1b but for CDM, $\Gamma = 0.35$.

**Fig. 6e** : Same as Fig.1a but for CDM, $\Gamma = 0.5$.

**Fig. 6f** : Same as Fig.1b but for CDM, $\Gamma = 0.5$.

**Fig. 6g** : Same as Fig.1a but for CDM, $\Gamma = 0.7$.

**Fig. 6h** : Same as Fig.1b but for CDM, $\Gamma = 0.7$.